# Parameter Estimation in Biokinetic Degradation Models in Wastewater Treatment - A Novel Approach Relevant for Micro-pollutant Removal

**Monika Schönerklee\*, Momtchil Peev\***

\* Austrian Research Centers GmbH - ARC, A – 2444 Seibersdorf, Austria (email: monika.schoenerklee@arcs.ac.at, momchil.peev@arcs.ac.at ), Tel.: +43 50 550 3450 (-3452 Fax)

## Abstract

In this paper we address a general parameter estimation methodology for an extended biokinetic degradation model [1] for poorly degradable micropollutants. In particular we concentrate on parameter estimation of the micropollutant degradation sub-model by specialised microorganisms. In this case we focus on the case when only substrate degradation data are available and prove the structural identifiability of the model. Further we consider the problem of practical identifiability and propose experimental and related numerical methods for unambiguous parameter estimation based on multiple substrate degradation curves with different initial concentrations. Finally by means of simulated pseudo-experiments we have found convincing indications that the proposed algorithm is stable and yields appropriate parameter estimates even in unfavourable regimes.

## Nomenclature

| Abbreviation | |
|---|---|
| X | biomass (µg COD l$^{-1}$) |
| S | substrate (µg COD l$^{-1}$) |
| µ | growth rate (d$^{-1}$) |
| µ$_{max}$ | maximum growth rate (d$^{-1}$) |
| K$_s$ | saturation constant (µg COD l$^{-1}$) |
| Y | growth yield |
| b | biomass decay rate (d$^{-1}$) |
| t | time (d) |
| | |
| Indices | |
| mp | micropollutant |
| 0 | Initial condition at t=0 |

**Introduction**

Quantitative mathematical modelling of wastewater treatment (as e.g. the well known IAWQ-ASM models) is a powerful tool allowing optimal design and operation of municipal treatment systems. However, standard models operate with integral system state-variables (such as COD) and do not enter into further details concerning specific (low-concentration) organic compounds with particular environmental impact. To achieve improved understanding and optimization of biological degradation of specific micropollutants, quantitative modelling going beyond the integral approach is required.

In this paper we address the parameter estimation for specific micropollutants to be included in an extended biokinetic degradation model [1]. It builds upon standard biokinetic model approaches, as e.g. ASM [2] and aims to describe the degradation of low-concentration organic substances by specialised minority bacterial sub-populations. To address such pollutants we concentrate on the extension of basic biokinetics with respect to a single process type – heterotrophic, aerobic biodegradation alone.

A major challenge for application of such extended models is parameter estimation, especially with respect to parameters related to the specialised biomass. The problem is to indirectly estimate its growth and decay rates, as to our best knowledge there are no easily available experimental methods for direct quantification of the concentration of specialised, active biomass. In this situation it has first of all to be clarified whether all parameters can be uniquely determined from substrate concentration measurements alone. To this end structural and practical identifiability analyses have to be performed.

**State of the art and motivation**

In a previous publication [1] we have addressed an extended biokinetic model describing the simultaneous biodegradation of two types of substrate, one of them being easily biodegradable ("normal") and a second one consisting of low-concentration micropollutants. The second substrate is assumed to be hardly biodegradable by a hypothetic specialised microorganism population only, which in reality can consist of many different species. This model contains three types of parameters:

- Parameters describing the "normal" biokinetic sub-system
- Parameters describing the specialised biokinetic sub-system
- Parameters describing the interactions of both sub-systems

This theoretical model extension has practical relevance only in case if the additional biokinetic model parameters are identifiable and can be determined by feasible experiments. The first type of parameters can easily be identified following standard parameter estimation methodologies [3, 4, 5]. The determination of the interaction parameters is currently an open question.

In this paper we focus specifically on determining the biokinetic parameters of the specialised sub-system. To this end we address the following experimental situation: Consider a biodegradation system where the full model described above is assumed to be relevant. By extracting a sludge sample and isolating it from sources of easily biodegradable ("normal") substrate, the specialised biokinetic sub-system can be considered as closed (see Equation 2 in [1]) and reduces to Equation (1). This reduced biokinetic sub-model describes biodegradation of micropollutant substrate by a specialised population only. It does not focus on particular substances, therefore in the model equations we have changed the indices $P^3$ (referring to persistent polar pollutants) in [1] to micropollutant (mp) compounds.

$$\left| \begin{aligned} \frac{dX_{mp}}{dt} &= \mu_{mp} X_{mp} - b_{mp} X_{mp} \\ \frac{dS_{mp}}{dt} &= -\frac{\mu_{mp}}{Y_{mp}} X_{mp} \end{aligned} \right. \quad (1)$$

Herebelow we restrict ourselves to standard Monod kinetics for the functional dependence of the specific growth rate $\mu_{mp}$:

$$\mu_{mp} = \mu_{max,mp} \frac{S_{mp}}{K_{s,mp} + S_{mp}} \quad (2)$$

We assume as usual that the kinetic parameters are constant and do not have a complex functional form.

As mentioned above in the situation at hand, due to the difficulty of direct measurements of active biomass $X_{mp}$, the parameter estimation methodology we have developed relies on substrate concentration measurements alone. Note that from an abstract point of view parameter estimation for biokinetic systems described by Equation (1) has long been studied under the assumption that both biomass and substrate concentrations (or other quantities which are functions of these variables) can be experimentally assessed [6]. The major difference of the approach we propose is based therefore on the assumption that only $S_{mp}$ measurements are available. (It should be noted that in case of micropollutants standard respirometric measurements are not feasible due to the extremely low substrate concentrations and the respective immeasurably weak respirometric signals.) As a consequence the methodology proposed here is not restricted to the case of micropollutants and can always be applied whenever Equation (1) is valid and only substrate concentration measurements are available.

There are two different questions to be answered [7]:

1. Is it possible at all to determine uniquely the biokinetic parameters even from ideal (error-free) experimental data? And if yes which parameters? This problem is known in literature as structural identifiability [8].

2. Is it possible to achieve unambiguous parameter estimation from real (noisy) experimental data? This problem is usually called practical identifiability [9].

**Structural Identifiability**

In this section we omit the subscripts $_{mp}$. This is done on the one hand for simplicity, but also because our results have a general mathematical validity (not restricted to micropollutants but applying to any biokinetic system described by equations of type (1, 2)).

From a mathematical point of view we have to solve the following problem: Given the (closed) system of equations (1, 2) and provided that the function S(t) is exactly known for all times (t ≥ 0), is it in principle possible to uniquely determine the set of biokinetic parameters $\mu_{max}$, $K_s$, $b$, $Y$ and the unknown initial condition (biomass at t=0) $X(0)=X_0$?

As a first step we note that by formal integration of the biomass equation one can transform the system of differential equations (1,2) into the following equivalent integro-differential equation.

$$\frac{dS}{dt} = -\mu_{max} \frac{X_0}{Y} \frac{S}{K_s + S} \exp\left\{\mu_{max} \int_0^t \frac{S}{K_s + S} d\tau - bt\right\} \quad (3)$$

This equation has one initial condition $S(0)=S_0$ and the other initial condition $X(0)=X_0$ appears as a new parameter. Already at this point it is clear that it is only the ratio $X_0/Y$ that can be structurally identifiable as any independent values of $X_0$ and $Y$ yielding the same ratio have an identical impact on equation (3). (This can be independently seen from equation (23) below.) Therefore the question is whether the set of parameters $\mu_{max}$, $K_s$, $b$ and $X_0/Y$ of equation (3) can be uniquely identified provided that the function $S(t)$ is known for all times.

To proceed further we take the logarithm of both sides of equation (3) multiplied by minus one.

$$\ln\left(-\frac{dS}{dt}\right) = \ln(\mu_{max}) + \ln\left(\frac{X_0}{Y}\right) + \ln\left(\frac{S}{K_s+S}\right) + \mu_{max}\int_0^t \frac{S}{K_s+S}d\tau - bt \quad (4)$$

Please note that this operation is well defined as both the right and the left hand side of equation (3) is a negative quantity for all $t \geq 0$.

Let us now consider a fixed solution of equation (3) $S(t)$, $t \geq 0$, $S(0)=S_0$. Then equation (4) is an algebraic relation between the set of four parameters $\mu_{max}$, $K_s$, $b$ and $X_0/Y$. Let us further assume that there exists a different set of parameters $\tilde{\mu}_{max}, \tilde{K}_s, \tilde{b}, \tilde{X}_0/\tilde{Y}$ which satisfies the following relation

$$\ln\left(-\frac{dS}{dt}\right) = \ln(\tilde{\mu}_{max}) + \ln\left(\frac{\tilde{X}_0}{\tilde{Y}}\right) + \ln\left(\frac{S}{\tilde{K}_s+S}\right) + \tilde{\mu}_{max}\int_0^t \frac{S}{\tilde{K}_s+S}d\tau - \tilde{b}t \quad (5)$$

for the same function $S(t)$, $t \geq 0$.

Subtracting the right hand sides of relations (4) and (5) we obtain the function $\alpha(t)$ which has to be identically equal to zero for all times $t \geq 0$ due to the fact that the left hand sides are identical.

$$\alpha(t) = \ln\left(\frac{\mu_{max}}{\tilde{\mu}_{max}}\right) + \ln\left(\frac{\frac{X_0}{Y}}{\frac{\tilde{X}_0}{\tilde{Y}}}\right) + \mu_{max}\int_0^t \frac{S(\tau)}{K_s+S(\tau)}d\tau - \tilde{\mu}_{max}\int_0^t \frac{S(\tau)}{\tilde{K}_s+S(\tau)}d\tau + \ln\left(\frac{\tilde{K}_s+S(t)}{K_s+S(t)}\right) + (b-\tilde{b})t \equiv 0, \quad \forall t \geq 0 \quad (6)$$

From equation (6) it immediately follows that

$$\alpha(0) = 0 \Rightarrow \frac{\tilde{X}_0}{\tilde{Y}} = \frac{\mu_{max}}{\tilde{\mu}_{max}} \frac{X_0}{Y} \left( \frac{\tilde{K}_s + S_0}{K_s + S_0} \right) \qquad (7)$$

Thus we have expressed the parameter $\tilde{X}_0/\tilde{Y}$ as a function of the remaining ones. Taking into account that the first derivative of equation (6) has also to be identically equal to zero for all times $t \geq 0$ and taking the limit $t \to \infty$ we find also that

$$\frac{d\alpha}{dt} = \mu_{max} \frac{S(t)}{K_s + S(t)} - \tilde{\mu}_{max} \frac{S(t)}{\tilde{K}_s + S(t)} - \mu_{max} \frac{X(t)}{Y} \frac{S(t)}{K_s + S(t)} \left( \frac{K_s - \tilde{K}_s}{(K_s + S(t))(\tilde{K}_s + S(t))} \right) + (b - \tilde{b}) \equiv 0, \quad \forall t \geq 0 \qquad (8)$$

$$\lim_{t \to \infty} \frac{d\alpha}{dt} = b - \tilde{b} \Rightarrow b = \tilde{b}$$

From the last result, the first derivative can be rewritten as follows:

$$\frac{d\alpha}{dt} = \frac{(\mu_{max} - \tilde{\mu}_{max})S(t)^3 + (\mu_{max}(K_s + \tilde{K}_s) - 2K_s\tilde{\mu}_{max})S(t)^2 + K_s(\mu_{max}\tilde{K}_s - \tilde{\mu}_{max}K_s)S(t) - \frac{\mu_{max}}{Y}(K_s - \tilde{K}_s)X(t)S(t)}{(K_s + S(t))^2(\tilde{K}_s + S(t))} = \frac{\varphi(t)}{\psi(t)} \equiv 0 \qquad (9)$$

where $\varphi(t) = (\mu_{max} - \tilde{\mu}_{max})S(t)^3 + (\mu_{max}(K_s + \tilde{K}_s) - 2K_s\tilde{\mu}_{max})S(t)^2 + K_s(\mu_{max}\tilde{K}_s - \tilde{\mu}_{max}K_s)S(t) - \frac{\mu_{max}}{Y}(K_s - \tilde{K}_s)X(t)S(t)$

and $\psi(t) = (K_s + S(t))^2(\tilde{K}_s + S(t))$

The function $\psi(t)$ is positive at all times and therefore the function $\Phi(t)$ has to be identically equal to zero. However, this function is a polynomial in the variables $S(t)$ and $X(t)$ (the span of which is a compact segment of the set of real positive numbers) and it can be identically equal to zero, if and only if the coefficients of the different powers are also identically equal to zero. This requirement automatically leads to

$$K_s = \tilde{K}_s \quad \text{and} \quad \mu_{max} = \tilde{\mu}_{max} \qquad (10)$$

These equalities are simultaneously necessary and sufficient conditions for the function $\Phi(t)$ to be identically equal to zero. Further from equations (10) and (7) it follows that:

$$\frac{\tilde{X}_0}{\tilde{Y}} = \frac{X_0}{Y} \qquad (11)$$

Equations (8), (10) and (11) demonstrate the unique structural identifiability of the parameters $\mu_{max}$, $K_s$, b and $X_0/Y$ of equation (3).

**Practical Identifiability**

Practical identifiability, i.e. determination of kinetic parameters from experimental data, is however a major challenge. In spite of the fact that it is possible to uniquely determine these parameters from error-free *single degradation curves*, as demonstrated above, this can hardly be done from experimental measurement data, the problems being mathematically ultimately related to the well known high sensitivity of biokinetic parameters encountered in standard microbial growth models [10, 11]. Therefore the potentially feasible methods to determine the biokinetic parameters would have to rely on additional information. One way to this end is to try to employ the information contained not in one but in series of degradation curves, for which it is on the one hand guaranteed that all relevant parameters $K_{s,mp}$, $\mu_{max,mp}$, the ratio $X_{0,mp}/Y_{mp}$ and $b_{mp}$ are identical while the initial values of the substrate concentrations $S_{0,mp}$ are different but known.

In this approach to solve the problem of practical identifiability there are two implicit tasks:

- To find a constructive method for determining the biokinetic parameters from ideal (error-free) experimental data of multiple degradation curves.
- To find a modus of application of this method to the case when the available experimental data are noisy and scarce.

The first of these two tasks can be solved explicitly. To this end we have developed an experimental design and protocol which is feasible for the case of micropollutants aiming at the estimation of the basic biokinetic parameters $K_{s,mp}$, $\mu_{max,mp}$, the ratio $X_{0,mp}/Y_{mp}$ and the decay rate $b_{mp}$ of the specialised biomass. We have designed two basic types of short-term batch experiments to determine these quantities. These should be thought of from two perspectives:

- First the experiments are sort of ideal "Gedankenexperimente" where it is assumed that they deliver sufficient data which are entirely free of noise. This allows to obtain a precise approximation by means of curve fitting of the decay and degradation curves.
- Second, the experiments are also suggested as real experiments by means of which the noisy and scarce experimental data are to be used to obtain reasonable approximation as described in detail below.

**Experimental design and constructive parameter estimation methodology in the ideal case**

1. Decay rate experiment - *Determination of the decay rate of specialised biomass $b_{mp}$ in mixed cultures*

The endogenous decay rate $b_{mp}$ is obtained by monitoring the decay of active cell mass in the absence of growth substrate. The major challenge in this case is finding a suitable method for evaluating the quantity of active biomass. A method that can be applied for mixed cultures, based on previous results by [12] can be described as follows:

The decay experiment is performed as a series of subsequent batch experiments using activated sludge in which effective biodegradation of the target micropollutant substance occurs. Initially a sludge sample is left to decay without substrate. From this decaying culture sub-samples are periodically withdrawn at time instances $t_i$. Each sub-sample is then spiked with a defined fixed concentration of micropollutant target substrate and the initial slope of each degradation curve is estimated. The procedure is based on the following relations:

- The initial slope of each of the resulting degradation curves $k_i$ (for $t=t_i$) is proportional to the concentration of active microorganisms $X_{mp}(t_i)$.

$$\frac{dS_{mp}}{dt} = -\frac{\mu_{mp}(S_{mp})}{Y_{mp}} X_{mp}(t_i) = k_i \qquad (12)$$

- The specialised (active) biomass fraction $X_{mp}(t_i)/X_{mp}(t_0)$ remaining at time $t_i$ for all i can be computed by dividing the initial slopes $k_i$ for each sub-sample by the initial slope $k_0$ ($t_0 = 0$) at the beginning of the decay period.

$$\frac{k_i}{k_0} = \frac{X_{mp}(t_i)}{X_{mp}(t_0)} = \frac{X_{0,mp} e^{-b_{mp} t_i}}{X_{0,mp}} = e^{-b_{mp} t_i} \qquad (13)$$

In Equation (12) the endogenous biomass decay process is described as well known by the function

$$X_{mp}(t_i) = X_{0,mp} e^{-b_{mp} t_i} \qquad (14)$$

which is a solution of Equation (1) in the case of absence of any substrate ($\mu = \mu_{mp} = 0$).

- The semilog plot of the specialised biomass fraction vs. time yields a straight line with slope of $-b_{mp}$.

Figure 1 shows the resulting idealised curve for the determination of the decay rate $b_{mp}$.

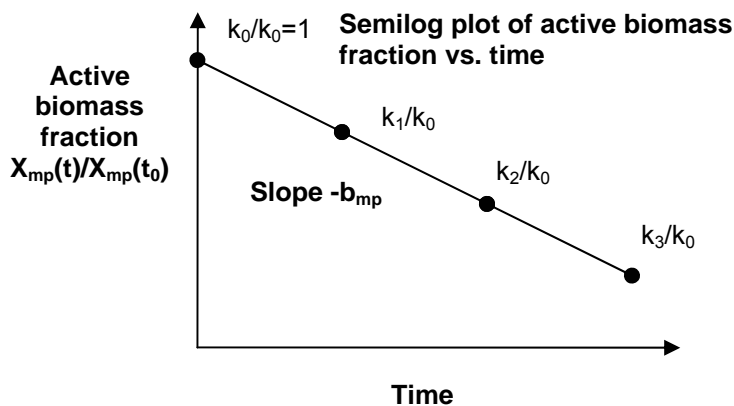

**Figure 1:** Determination of the specialised biomass decay rate

2. Degradation experiments *for determination of the biokinetic constants $K_{s,mp}$, $\mu_{max,mp}$, $X_{0,mp}/Y_{mp}$ and $b_{mp}$*

This experiment is performed as a series of parallel batch experiments using identical activated sludge samples in which effective biodegradation of the target micropollutant substance occurs. In each sample a chosen concentration of the target micropollutant substrate is spiked, whereby the concentrations for the separate samples are different and form a monotonous series. The basic procedure includes the following steps:

- Monitoring the degradation of selected micropollutant substrate, caused by specialised biomass. For different initial substrate concentrations the resulting degradation curves are recorded for m+1 different time instances (0, $t_1$, …, $t_m$). For n different initial substrate concentrations $S_{1,mp}(0), \ldots, S_{n,mp}(0)$ these curves can be seen as solutions of the following differential equations:

$$\frac{dS_{1,mp}}{dt} = -\frac{\mu_{max,mp}}{Y_{mp}} \frac{S_{1,mp}(t)}{K_{s,mp} + S_{1,mp}(t)} X_{1,mp}(t),$$
…
$$\frac{dS_{n,mp}}{dt} = -\frac{\mu_{max,mp}}{Y_{mp}} \frac{S_{n,mp}(t)}{K_{s,mp} + S_{n,mp}(t)} X_{n,mp}(t).$$

(15)

- Determination of $K_{s,mp}$, $\mu_{max,mp}$ and $X_{0,mp}/Y_{mp}$

    For t=0 these differential equations can be seen as a set of algebraic equations of the form:

$$\frac{dS_{1,mp}}{dt}(0) = -\frac{\mu_{max,mp}}{Y_{mp}} \frac{S_{1,mp}(0)}{K_{s,mp} + S_{1,mp}(0)} X_{1,mp}(0)$$
…
$$\frac{dS_{n,mp}}{dt}(0) = -\frac{\mu_{max,mp}}{Y_{mp}} \frac{S_{n,mp}(0)}{K_{s,mp} + S_{n,mp}(0)} X_{n,mp}(0)$$

(16)

Note that due to the experiment design $X_{1,mp}(0)=\ldots= X_{n,mp}(0)$. In what follows we denote this parameter with $X_{0,mp}$. Equations (16) are linear with respect to the two unknown variables ($x_1=K_{s,mp}$ and $x_2=\mu_{max,mp}*X_{0,mp}/Y_{mp}$).

$$\hat{A}\vec{x}=\vec{y}, \quad \hat{A}=\begin{pmatrix} \frac{dS_{1,mp}}{dt}(0) & S_{1,mp}(0) \\ \ldots & \ldots \\ \ldots & \ldots \\ \ldots & \ldots \\ \frac{dS_{n,mp}}{dt}(0) & S_{n,mp}(0) \end{pmatrix}, \quad \vec{x}=\begin{pmatrix} x_1=K_{s,mp} \\ x_2=\mu_{max,mp}\frac{X_{0,mp}}{Y_{mp}} \end{pmatrix}, \quad \vec{y}=\begin{pmatrix} -S_{1,mp}(0)\frac{dS_{1,mp}}{dt}(0) \\ \ldots \\ \ldots \\ \ldots \\ -S_{n,mp}(0)\frac{dS_{n,mp}}{dt}(0) \end{pmatrix} \quad (17)$$

As it is well known [13], provided that the columns of $\hat{A}$ are linearly independent then exists

$$\tilde{\vec{x}}=\hat{A}^+\vec{y}, \quad \text{where} \quad \hat{A}^+=(\hat{A}^T\hat{A})^{-1}\hat{A}^T \quad \text{is the pseudoinverse matrix of } \hat{A}. \quad (18)$$

The pseudoinverse provides a least squares solution to a system of linear equations. Thus $\tilde{\vec{x}}$ is equal to the least square solution of the system given in equation (17).

Abstractly, the linear independence of the columns of $\hat{A}$ is always guaranteed. Indeed, as

$$\hat{A}=\begin{pmatrix} \frac{dS_{1,mp}}{dt}(0) & S_{1,mp}(0) \\ \ldots & \ldots \\ \ldots & \ldots \\ \ldots & \ldots \\ \frac{dS_{n,mp}}{dt}(0) & S_{n,mp}(0) \end{pmatrix} = -\mu_{max.mp}\frac{X_{0,mp}}{Y_{mp}}\begin{pmatrix} \frac{S_{1,mp}(0)}{K_{s,mp}+S_{1,mp}(0)} & S_{1,mp}(0) \\ \ldots & \ldots \\ \ldots & \ldots \\ \ldots & \ldots \\ \frac{S_{n,mp}(0)}{K_{s,mp}+S_{n,mp}(0)} & S_{n,mp}(0) \end{pmatrix} \quad (19)$$

it is only in the regime when $K_{s,mp} \gg S_{i,mp}(0)$ (i=1,…,n) that $\hat{A}$ has effectively rank equal to one and taking the pseudoinverse is practically impossible.

- The determination of $\mu_{max,mp}$ plus independent estimations of $\mu_{max,mp}*X_{0,mp}/Y_{mp}$ and $b_{mp}$ is based on equation (4) applied to all experimental data for all degradation curves and is carried out as follows:

$$\hat{B}\,\vec{z} = \vec{v} \tag{20}$$

$$\hat{B} = \begin{pmatrix} 1 & -t_1 & \int_0^{t_1} \frac{S_{1,mp}}{K_{s,mp}+S_{1,mp}} d\tau \\ \ldots & \ldots & \ldots \\ 1 & -t_1 & \int_0^{t_1} \frac{S_{n,mp}}{K_{s,mp}+S_{n,mp}} d\tau \\ 1 & -t_2 & \int_0^{t_2} \frac{S_{1,mp}}{K_{s,mp}+S_{1,mp}} d\tau \\ \ldots & \ldots & \ldots \\ 1 & -t_m & \int_0^{t_m} \frac{S_{n,mp}}{K_{s,mp}+S_{n,mp}} d\tau \end{pmatrix}, \vec{z} = \begin{pmatrix} z_1 = \ln\left(\mu_{max,mp} \frac{X_{0,mp}}{Y_{mp}}\right) \\ z_2 = b_{mp} \\ z_3 = \mu_{max,mp} \end{pmatrix}, \vec{v} = \begin{pmatrix} \ln\left(-\frac{dS_{1,mp}}{dt}(t_1)\right) - \ln\left(\frac{S_{1,mp}(t_1)}{K_{s,mp}+S_{1,mp}(t_1)}\right) \\ \ldots \\ \ln\left(-\frac{dS_{n,mp}}{dt}(t_1)\right) - \ln\left(\frac{S_{n,mp}(t_1)}{K_{s,mp}+S_{n,mp}(t_1)}\right) \\ \ln\left(-\frac{dS_{1,mp}}{dt}(t_2)\right) - \ln\left(\frac{S_{1,mp}(t_2)}{K_{s,mp}+S_{1,mp}(t_2)}\right) \\ \ldots \\ \ln\left(-\frac{dS_{n,mp}}{dt}(t_m)\right) - \ln\left(\frac{S_{n,mp}(t_m)}{K_{s,mp}+S_{n,mp}(t_m)}\right) \end{pmatrix} \tag{21}$$

$\tilde{\vec{z}} = \hat{B}^+\,\vec{v}$, where $\hat{B}^+ = (\hat{B}^T\hat{B})^{-1}\hat{B}^T$ is the pseudoinverse matrix of $\hat{B}$. (22)

Once again, the linear independence of the columns of $\hat{B}$ is abstractly always given, but this time in the regime when $K_{s,mp} \ll S_{i,mp}(0)$ (i=1,…,n) $\hat{B}$ has effectively rank equal to two and taking the pseudoinverse is practically impossible.

In summary, we have identified methods to constructively determine $K_{s,mp}$, $\mu_{max,mp}$, $X_{0,mp}/Y_{mp}$ and $b_{mp}$ from ideal decay and multiple degradation experiments. Two quantities, namely $b_{mp}$ and $\mu_{max,mp}*X_{0,mp}/Y_{mp}$ are independently determined in two different ways. It should be stressed however that the second of these quantities, i.e. $\mu_{max,mp,}*X_{0,mp}/Y_{mp}$ is determined using two different algorithms from the *same* experimental data. In contrast the decay rate is determined from *two independent experiments*. Thus abstractly Experiment 1, above can be skipped. On the other hand it increases the confidence of the results of Experiment 2.

**Parameter estimation from realistic experimental data**

To go to realistic (noisy) experiments, one has to be able to obtain good curve fits for each of the degradation curves and at least the initial slope of the decay curves. A straightforward way to do so is to use some standard curve-fit methodology. Unfortunately for few experimental points and some degree of noise the result can be strongly deviating if some standard family of interpolation functions as e.g. splines is applied. A feasible approach can be defined as follows:

- Take some reasonable, but imprecise approximation of the degradation curves by some family of fitting functions which most appropriately have to monotonously decrease as a function of time as do the degradation curves.

- This allows the initial determination of the biokinetic parameters: first of all $b_{mp}$, then $K_{s,mp}$ and the product $\mu_{max,mp} \ast X_{0,mp}/Y_{mp}$ and finally $\mu_{max,mp}$ plus an independent estimation of $b_{mp}$ and $\mu_{max,mp} \ast X_{0,mp}/Y_{mp}$.

- Now one can attempt a renewed approximation of the experimental data, this time by numerical solution of the system of differential equations (1) and (2) by using the biokinetic parameters estimated in the previous step and least square fitting. Additionally this approach takes advantage of the fact that

$$\left| \begin{array}{l} \dfrac{dX_{mp}}{dt} = \mu_{mp} X_{mp} - b_{mp} X_{mp} \\ \dfrac{dS_{mp}}{dt} = -\dfrac{\mu_{mp}}{Y_{mp}} X_{mp} \end{array} \right. \Leftrightarrow \left| \begin{array}{l} \dfrac{dx_{mp}}{dt} = \mu_{mp} x_{mp} - b_{mp} x_{mp} \\ \dfrac{dS_{mp}}{dt} = -\mu_{mp} x_{mp} \end{array} \right. \qquad (23)$$

$$S_{mp}(0) = S_{0,mp},\ X_{mp}(0) = X_{0,mp} \qquad S_{mp}(0) = S_{0,mp},\ x_{mp}(0) = \dfrac{X_{0,mp}}{Y_{mp}}$$

This allows to find solutions for $S_{mp}$ although we cannot determine $Y_{mp}$ separately.

- Finally one can once again determine the biokinetic parameters making use of these new approximations of the degradation curves and iterate all algorithms until stable results are achieved.

The above discussion does not yet allow estimating the confidence region of the evaluated biokinetic parameters in case of noisy data and scarce experimental points. While a full statistical analysis of the proposed method is still missing, below we give an initial "experimental" estimate of these confidence regions by means of simulating pseudo-experimental data.

**"Noise simulation"**

The pseudo-experimental data are obtained by taking calculated ideal values and adding statistically independent noise. We have deliberately chosen to have a small number of measurement points as in real experiments only a limited number of analysed samples is usually available. This is due to a number of reasons: For low-concentration micropollutants the analysis procedure is extremely complex, time-consuming and costly. Additionally sample collection under identical experimental conditions is prerequisite for sound parameter estimation. While this is still possible for a small number of samples, it becomes increasingly difficult with larger sample series.

In particular the pseudo-experimental data are obtained as follows:

- We simulate equation (23) for the following parameter values: $\mu_{max,mp}$ =1 d$^{-1}$, $K_{s,mp}$= 22 µg/l, $X_{0,mp}/Y_{mp}$=330 µg/l, $b_{mp}$=0,3 d$^{-1}$ for four fixed initial values of the micropollutant substrate concentration at t=0.
- We add normally distributed noise (four values of standard deviations stabw= 0, 0.02, 0.05, 0.1) for each pseudo-experimental value. Adding noise technically means that we produce an array of normally distributed random numbers with a mean value of 1 (the length of the array being equal to the length of the pseudo-experimental series) and multiply in a point-by-point way the random number array and the pseudo-experimental data array.

- For each value of standard deviation of the noise we produce the data for a number of noisy pseudo-experiments as described above.
- From each of these pseudo-experimental noisy curves we take out a sub-set consisting of 7 "experimental points" at fixed time instances.

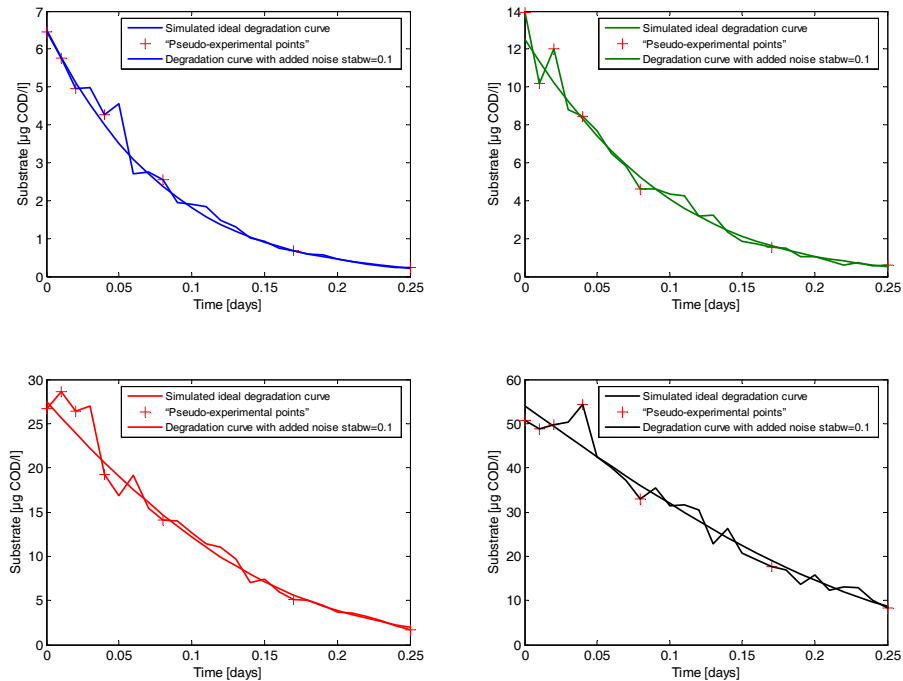

**Figure 2:** Plot of a set of simulated ideal degradation curves, these degradation curves with added noise stabw=0.1 and "experimental points" at seven fixed time instances

One should note that these final pseudo-experimental series are really unfavourable for parameter estimation. In addition to the mentioned scarcity of data points, the noise we have introduced is proportional to the "true" pseudo-experimental values. This needs not actually be the case in a real measurement error as the distribution could be constrained in a fixed range and not necessarily proportional to the experimental value. With the approach we have chosen we particularly increase the noisiness of degradation curves with higher initial substrate concentrations and in general the "error" in the initial part of every degradation curve. This affects particularly these parts of our estimation algorithm which are based on initial slopes and substrate concentrations.

For each of the pseudo-experimental series we apply the algorithmic approach for parameter estimation as described above using 100 iteration steps. The results are arrays of parameter estimates indexed by the iteration step of the algorithm. An example of such type of results for standard deviation 0.02 is presented in Figure 3.

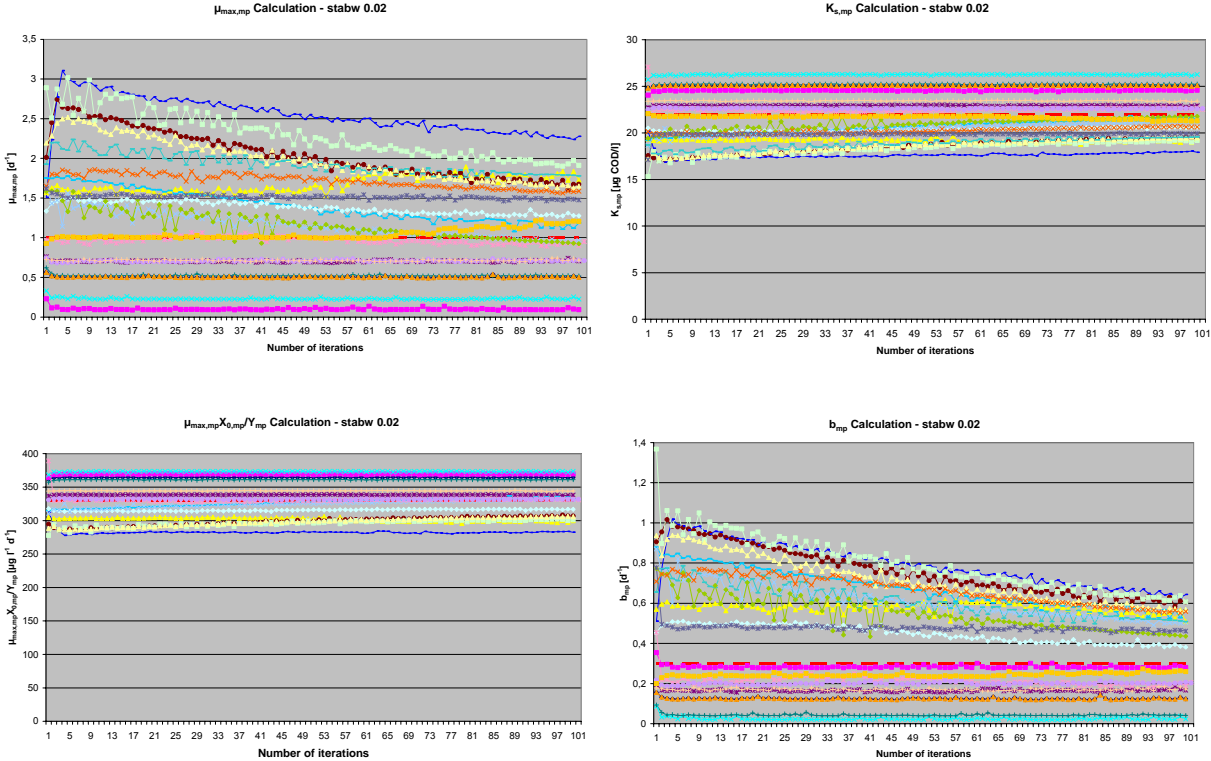

**Figure 3:** Calculated values for $\mu_{max,mp}$, $K_{s,mp}$, $\mu_{max,mp} * X_{0,mp}/Y_{mp}$ and $b_{mp}$ using noise standard deviation stabw=0.02 for 22 sets of "pseudo-experimental data" using 100 iterations in each parameter estimation

The results of all parameter estimations based on pseudo-experimental data are given in Figure 4, whereby the straight line corresponds to the values of the parameters used in the simulation before adding noise. (Note that these values coincide with the algorithm output in the case of no noise – standard deviation of noise equal to zero). For each value of the "input noise" standard deviation a mean value of all algorithm estimates and the corresponding "output" standard deviation are given.

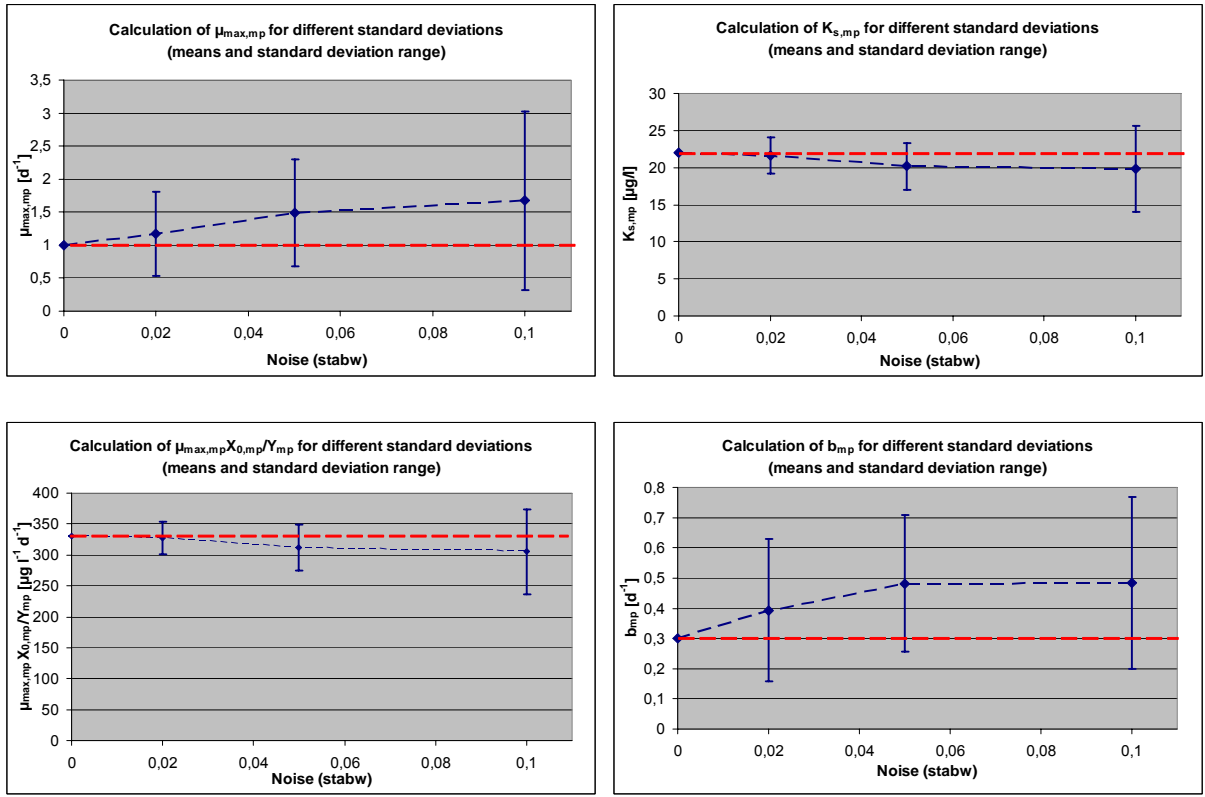

**Figure 4:** Diagramme for $\mu_{max,\,mp}$, $K_{s,mp}$, $\mu_{max,mp} *X_{0,mp}/Y_{mp}$ and $b_{mp}$ for different standard deviations (mean plus confidence range) for 22 sets of "pseudo-experimental data" using 100 iterations in each parameter estimation

This approach gives satisfactory indications of the standard deviation range of the evaluated biokinetic parameters as a function of the degree of "noisiness" characterised by e.g. the standard deviation of the applied noise function. Simultaneously it should be noted that it cannot account for any modelling errors – i.e. possible deviations of the experimental data to the form of the model assumed.

**Conclusions**

In this paper we have addressed the structural and practical identifiability of equations (1) and (2) in the case when only substrate measurement data are available from a methodological point of view. We have proven the structural identifiability of the problem and have developed a new approach allowing practical identifiability on the basis of multiple substrate degradation curves with different initial concentrations. While a full statistical sensitivity analysis is still missing, we have shown by means of simulated pseudo-experimental noisy data that the proposed parameter estimation algorithm is stable and gives appropriate parameter estimates even in a strongly unfavourable regime chosen above.

The approach described above is the basis for parameter estimation for a number of micropollutant substances investigated in activated sludge and MBR systems. In particular, specific experiments were carried out in order to determine biokinetic degradation parameters for NDSA and BTSA for which the presented methodology was additionally tested and verified. The experimental results and the corresponding parameter estimation will be presented in a subsequent publication.

**Acknowledgement**

The presented work was carried out within the framework of the project PTHREE (www.pthree.de) financed under contract number EVK1-CT-2002-00116 by the European Commission (FP5).